\documentclass[prb,twocolumn,showpacs]{revtex4}
\usepackage{graphicx}
\usepackage{amsmath}
\usepackage{amssymb}

\renewcommand{\Im}{\mathrm{Im}\,}

\renewcommand{\Im}{\mathrm{Im}\,}

\newcommand{\ra}{\rangle}
\newcommand{\la}{\langle}
\DeclareMathAlphabet{\bi}{OML}{cmm}{b}{it}
\begin{document}
\title{Zero-field spin splitting in a two-dimensional electron gas
with the spin-orbit interaction revisited}

\author{SK Firoz Islam and Tarun Kanti Ghosh}
\affiliation{Department of Physics, Indian Institute of Technology-Kanpur,
Kanpur-208 016, India}

\begin{abstract}
We consider a two-dimensional electron gas (2DEG) with the Rashba 
spin-orbit interaction (SOI) in presence of a perpendicular magnetic 
field. We derive analytical expressions of the density of states (DOS) 
of a 2DEG with the Rashba SOI in presence of magnetic field by using 
the Green's function technique. The DOS allows us to obtain the analytical 
expressions of the magnetoconductivities for spin-up and spin-down electrons.
The conductivities for spin-up and spin-down electrons oscillate with 
different frequencies and gives rise to the beating patterns in the amplitude 
of the Shubnikov de Hass (SdH) oscillations. We find a simple equation 
which determines the zero-field spin splitting energy if the magnetic field 
corresponding to any beat node is known from the experiment.
Our analytical results reproduce well the experimentally observed 
non-periodic beating patterns, number of oscillations between 
two successive nodes and the measured zero-field spin splitting energy.

\end{abstract}


% 71.70.Ej, Spin-orbit coupling in condensed matter
% Spintronics, 85.75.-d
% spin transport effects, 75.76.+j
%73.43.Qt magnetoresistance
%72.10-d theory of electric transport and scattering mechanism

%73.20. At surface states, band structure electron DOS

\date{\today}

\maketitle
\section{INTRODUCTION}
There has been tremendous research interest on the spin-orbit interaction (SOI) 
in low-dimensional condensed matter systems due to the possible applications
in spin based electronic devices \cite{rmp1,rmp2,book}.
The SOI is responsible for many noble effects like spin-FET \cite{das1,bandyo}, 
metal-insulator transition in a two-dimensional hole gas \cite{mit}, 
spin-resolved ballistic transport \cite{ballistic}, spin-galvanic effect \cite{galvanic} 
and spin Hall effect \cite{perel,she}. It is the prime interest to
measure and control the SOI strength experimentally because it influences
the spin degree of freedom.

In absence of an external magnetic field, the two-fold spin degeneracy is lifted 
at finite momentum of the electron due to the SOI.
In semiconductor heterostructures, there are mainly two mechanisms responsible 
for giving rise to the zero-field spin splitting: the Dresselhaus
interaction\cite{dress} which varies as $k^3$  and 
the Rashba interaction \cite{rashba} which varies linear with $k$ .
The former is due to the inversion asymmetry of the crystals and the
later is due to the asymmetric quantum wells.
The Dresselhaus interaction dominates in the wide-gap semiconductors 
with small thickness whereas the Rashba interaction dominates in narrow-gap semiconductors 
due to their different momentum dependence \cite{lommer,bassani}.
The electric field generated in the asymmetric quantum wells produces the Rashba
spin-orbit interaction. The SOI can also be tuned by a strong external electric
field perpendicular to the planar motion of the electrons. 

A direct manifestation of the spin-split levels due to the SOI is a beating
pattern in the SdH oscillations due to the two closely spaced different frequency 
of the spin-up and spin-down electrons. 
In Ref. \cite{bassani}, it was shown that the Rashba interaction produces regular beating patterns
in the SdH oscillations whereas the Dresselhaus interaction produces anomalous beating patterns.
The SOI strength is measured by analyzing the beating patterns in the SdH oscillations.

The experimental evidence, by using the magnetotransport and 
cyclotron resonance measurement, of zero-field spin splitting is found 
in a modulation doped GaAs/AlGaAs heterojunction \cite{stormer,stein}. 
It was first explained by Bychkov and Rashba based on the spin-orbit
interaction produced in the asymmetric quantum wells \cite{rashba1}. 
The zero-field spin splitting is related to the Fermi wave vector
$k_{_F} $ and the SOI strength $\alpha $ as
$ \Delta_s = 2 k_{_F} \alpha $. 
The Rashba SOI is considered to be the appropriate term for observing the
zero-field spin splitting in low-dimensional quantum systems, particularly in 
narrow gap semiconductors. In Ref. \cite{luo}, the beating patterns in the SdH 
oscillations have been found in GaSb/InAs quantum wells and confirmed that 
the lifting of the spin degeneracy is due to the Rashba SOI. 
Later, Das et al. \cite{dutta} investigated the SdH oscillations in 
a series of three different modulation-doped heterostructures with
high electron densities and confirmed that the lifting of the spin
degeneracy is due to the Rashba SOI.
There are many advanced technique to control the SOI strength in 2DEG
of different materials by varying the gate voltage \cite{tech,tech1,tech2}.

In Ref. \cite{das}, the zero-field spin splitting energy and hence 
the Rashba SOI strength were determined by extrapolating data and by using 
the model calculations for the SdH oscillations. 
Later, the zero-field spin splitting energy was studied theoretically
based on the difference between the Landau energy levels \cite{wang} and
the self-consistent Born approximation \cite{other}.  
The estimated SOI strength was in good agreement with the extrapolated
results obtained in \cite{dutta}.

In this work, we provide a very simple and elegant way to determine the 
zero-field spin splitting energy. The analytical expressions 
of the SdH oscillations for the spin-up and spin-down electrons are also obtained. 
The total magnetoconductivity shows beating patterns in the SdH oscillations
due to two closely spaced different frequencies of the SdH oscillations 
for spin-up and spin-down electrons.   
By analyzing the beating patterns, we find a very simple equation to determine the
zero-field spin splitting energy from the location of any beat node.
We also explain analytically the non-periodic behavior of the beating patterns and
the number of oscillations between two successive nodes.

This paper is organized as follows. In section II, we summarize the 
exact energy eigenvalues and the corresponding eigenfunctions of a 2DEG 
with the Rashba SOI in presence of a perpendicular constant magnetic field.
We calculate the DOS of a 2DEG with the SOI in presence of a magnetic field.
Numerical and analytical results of the SdH oscillations are presented in section III. 
In section V, we use available experimental data to calculate the zero-field 
spin splitting energy, number of oscillations between two nodes from our analytical results. 
We present a summary of our work in Sec. VI.

\section{ENERGY SPECTRUM AND DENSITY OF STATES IN PRESENCE OF RASHBA SOI}
The Hamiltonian of an electron $(-e)$ with the Rashba SOI 
in presence of a perpendicular magnetic field ${\bf B} = B \hat z$ is given by
\begin{equation}
H = \frac{({\bf p} + e {\bf A})^2}{2m^{\ast }} \mathbb{1} + 
\frac{\alpha}{\hbar }\left[ {\mbox{\boldmath $\sigma$} } \times 
({\bf p}+e{\bf A})\right]_z + \frac{1}{2}g\mu _{_B} B \sigma_z, \label{Ham}
\end{equation}
where ${\bf p}$ is the 2D momentum operator, $m^{\ast }$ is the effective mass 
of the electron, $g$ is the Lande-g factor, $ \mu_{_B} $ is the Bohr magneton,
$\mathbb{1} $ is the unit matrix, 
${\mbox{\boldmath $\sigma$} }=(\sigma_x,\sigma_y,\sigma_z)$ are the Pauli 
spin matrices, and $\alpha $ is the strength of the Rashba interaction.

Here, we shall just mention the exact solutions of 
the Hamiltonian $H$. Using the Landau wave functions without 
the Rashba SOI as the basis, one can obtain the energy 
spectrum and the corresponding eigenfunctions \cite{wang}. 
The resulting eigenstates are labeled by a new quantum number $s$, 
instead of the Landau level quantum number $n$ in absence of the SOI.
For $s=0$  there is only one level, the same as the lowest Landau
level without SOI, with energy
$ E_0^+ = E_0 = (\hbar \omega - g \mu_{_B} B)/2 $
and the corresponding wave function is
\begin{equation}
\Psi_0^+ (k_y)= \frac{e^{i k_y y}}{\sqrt{L_y}} \phi_0(x + x_0)
\left(
\begin{array}{c}
0 \\
1
\end{array}
\right).
\end{equation}
Here, $ \omega = eB/m^* $, $x_0 = k_y l_0^2$ with 
$l_0 = \sqrt{\hbar/(eB)}$ is the magnetic length scale and
$ \phi_0 (x) = e^{-x^2/(2l_0^2)}/\sqrt{\sqrt{\pi}l_0}$.
For $s=1,2,3....$ there are two branches of the energy levels, 
denoted by $+$ corresponding to the "spin-up" electrons 
and $-$ corresponding to the "spin-down" electrons with energies
\begin{equation}
E_s^{\pm} = s\hbar\omega {\pm} \sqrt{E_0^2 + s E_{\alpha} \hbar \omega},
\end{equation}
where $ E_{\alpha} = 2 m^{\ast} \alpha^2/\hbar^2$ is the Rashba energy determined
by the SOI strength $ \alpha $.
The corresponding wave function for $+$ branch is
\begin{equation}
\Psi _s^+ (k_y)=\frac{e^{i k_y y}}{\sqrt{L_y A_s }}\left(
\begin{array}{r}
D_s \phi_{s-1}(x + x_0)
\\
\phi_s (x + x_0)
\end{array}
\right) \text{,}
\end{equation}
and the $-$ branch is
\begin{equation}
\Psi_s^-(k_y)=\frac{e^{i k_y y}}{\sqrt{L_y A_s }}\left(
\begin{array}{r}
\phi_{s-1}(x + x_0)
\\
- D_s \phi_s(x + x_0)
\end{array}
\right) \text{,}
\end{equation}\\
where $A_s = 1 + D_s^2 $,
$ D_s = \sqrt{s E_{\alpha} \hbar \omega}/
[E_0 + \sqrt{E_0^2 + s E_{\alpha} \hbar \omega}]$.
Also,
$ \phi_s(x) = e^{-x^2/(2l_0^2)} H_s(x/l_0)/\sqrt{\sqrt{\pi} 2^s s! l_0} $ 
is the harmonic oscillator wavefunctions and $H_s(x) $ is the Hermite
polynomial.

We calculate the density of states (DOS) $D(E)$ of a 2DEG 
with the Rashba SOI in presence of a weak magnetic field. 
This is calculated by taking imaginary part of the electron's 
self-energy $ \Sigma^{-}(E)$ using the expression 
$ D(E) = \Im \{\Sigma^{-}(E)/(\pi^2l_0^2\Gamma_0^2)\}$. 
Here, $\Gamma_0 $ is the impurity induced Landau level broadening.
The DOS for spin-up and spin-down electrons are calculated and 
these are given as 
\begin{eqnarray}
D^{\pm}(E) & = & \frac{m^*}{2\pi \hbar^2} 
\Big[1 + 2 \exp{\Big\{-2\Big(\frac{\pi\Gamma_0}{\hbar\omega}\Big)^2\Big\}} 
\nonumber \\ 
& \times & \cos{\Big\{\frac{2\pi}{\hbar\omega}
\Big(E+\frac{E_{\alpha}}{2}\mp \sqrt{E_0^2+E_{\alpha}E}\Big)\Big\}} \Big] \label{dos}.
\end{eqnarray}
The analytical expressions of the DOS given in Eq. (\ref{dos}) will be used to 
calculate the magnetoconductivity in the next section.

In absence of the Rashba and the Zeeman terms, the above mentioned 
DOS reduces to the well-known result of the DOS of a 2DEG without 
SOI in presence of the magnetic field \cite{ando,ger}:
\begin{eqnarray}
D(E)  =  \frac{m^*}{\pi \hbar^2}
\Big[ 1 - 2 \exp \Big \{ - 2 \Big( \frac{\pi \Gamma_0}{\hbar \omega} \Big)^2 \Big \}  
\cos \Big ( \frac{2\pi E}{\hbar\omega} \Big ) \Big].
\end{eqnarray}

\section{Calculation of Magnetoconductivity}
In general, there are two scattering 
mechanisms, diffusive and collisional scattering, 
contribute to the transport properties. 
The diffusive scattering is due to 
finite drift velocity gained by the electrons. 
In our case there is no finite group velocity along
$y$-direction due to the $k_y$ degeneracy in the energy spectrum. 
Therefore the diffusive contribution to the total conductivity is zero. 
The collisional conductivity 
arises because of the migration of the cyclotron orbit due
to scattering from the charge impurities present in the system.
In our problem, the diagonal conductivity 
$ \sigma_{xx} = \sigma_{xx}^{\rm col} $ because 
$\sigma_{xx}^{\rm dif} = \sigma_{yy}^{\rm dif} = 0$. 
The magnetoresistivity is 
$\rho_{yy} = \sigma_{xx}/S $, where 
$S = \sigma_{xx} \sigma_{yy} - \sigma_{xy} \sigma_{yx} 
= \sigma_{xy}^2 $ with $ \sigma_{xy} \simeq n_e e/B$.

At low temperature, we assume that electrons are scattered 
elastically by charged impurities distributed uniformly. The 
standard expression for collisional conductivity \cite{kubo} 
is given by
\begin{equation}
\sigma_{\mu \mu}^{col} = \frac{\beta e^2}{S_0}
\sum_{\xi, \xi^{\prime}} f_{\xi} (1-f_{\xi}) 
W_{\xi, \xi^{\prime}} 
(\alpha_\mu^{\xi} - \alpha_{\mu}^{\xi^{\prime}})^2,
\end{equation}
where $ W_{\xi,\xi^{\prime}} $ is the transition between 
one-electron states $ |\xi \ra $ and $ | \xi^{\prime} \ra $. 
Also, $ \alpha_{\mu}^{\xi} = \la \xi | r_{\mu} | \xi \ra $ is
the expectation value of the $\mu$ component of the position 
operator when the electron is in the the state $ |\xi \ra $.
The scattering rate $ W_{\xi,\xi^{\prime}} $ is given by
\begin{equation}
W_{\xi,\xi^{\prime}} = \sum_{\bf q_{_0}} | U ({\bf q_{_0}})|^2 \la \xi | 
e^{i {\bf q_{_0}} \cdot ({\bf r } - {\bf R})}| \xi^{\prime} \ra|^2 
\delta(E_{\xi} - E_{\xi^{\prime}}),    
\end{equation}
where $ {\bf q_{_0}} = q_{_{0x}} \hat x + q_{_{0y}} \hat y$ is 
the 2D wave-vector and 
$ U({\bf q_{_0}}) = 2 \pi e^2/
(\epsilon \sqrt{q_{_{0x}}^2 + q_{_{0y}}^2 + k_s^2}) $ 
is the Fourier transform of the screened impurity potential 
$ U({\bf r}) = (e^2/4\pi \epsilon) (e^{-k_sr}/r) $, where
$k_s$ is the inverse screening length and $ \epsilon $ is 
the dielectric constant of the material. 
In the limit of small $|{\bf q_{_0}}| \ll k_s $, 
$ U({\bf q_{_0}}) \simeq 2\pi e^2/(\epsilon k_s) = U_0$.
Here, ${\bf r} $ and ${\bf R} $ are the position vector of 
electron and impurity, respectively.
Finally the conductivity for spin-up and spin-down electrons 
become \cite{wang}
\begin{equation} 
\sigma_{xx}^{\pm} = \frac{e^2}{h}\frac{ \beta N_I U_0^2}{2\pi \Gamma_0 l_0^2} 
\sum_{s} I_s^{\pm} f_s^{\pm}(1-f_s^{\pm}) \label{exact},
\end{equation}
where $N_I$ is the 2D impurity number density and 
$f_s^{\pm}$ is the Fermi-Dirac distribution function.
The exact expressions of $ I_s^{\pm}$ are given as
$ I_s^{\pm} = [(2s\mp1)D_s^4-2sD_s^2+(2s\pm1)]/A_s^2$.

We would like to derive analytical expressions of the conductivities
for spin-up and spin-down electrons 
by using the DOS given in Eq. (\ref{dos}). 
The summation over quantum number $s$ in Eq. (\ref{exact})
can be replaced as $\sum_s \rightarrow  2 \pi l_0^2 \int_{0}^{\infty}D(E) dE$.
After a lengthy calculation, we obtain the analytical expressions of 
the conductivities for spin-up and spin-down electrons, which are given by
\begin{eqnarray}
\frac{\sigma_{xx}^{\pm}}{\sigma_0} & = &  
\frac{ \tilde E_F}{8(\omega\tau)^2} 
\Big[1 + 2\exp{\Big\{-2\Big(\frac{\pi\Gamma_0}
{\hbar\omega}\Big)^2\Big\}} \nonumber \\ 
& \times &
A\Big(\frac{T}{T_c}\Big)
\cos{\Big(\frac{2\pi f^{\pm}}{B}\Big)}\Big],\nonumber\\
\end{eqnarray}
where $\sigma_0 = n_e e^2 \tau/m^*$ is the Drude conductivity,
$\tilde{E}_{F} = \Big[1 + \frac{E_{\alpha}}{2E_{F}} \mp
\frac{3}{2}\sqrt{\frac{E_{\alpha}}{E_{F}}}\Big]$,
$ A\Big(T/T_c\Big)= (T/T_c)/\sinh(T/T_c)$ with
$T_c = \hbar \omega/2\pi^2k_{_B}$.
The conductivities for spin-up and spin-down electrons are oscillating
with different frequencies $f^{\pm} $ (in Tesla) as given below:
\begin{equation}
f^{\pm} = \frac{m^*}{\hbar e} \Big[E_{F} + 
\frac{E_{\alpha}}{2} \mp \sqrt{E_0^2 + E_{\alpha} E_{F}} \Big].
\end{equation}
The total conductivity is given by
\begin{eqnarray} 
\frac{\sigma_{xx}}{\sigma_0} &  \simeq & 
\frac{1}{4(\omega \tau)^2} \Big[1 +  2 \exp{\Big\{-2\Big(\frac{\pi\Gamma_0}
{\hbar\omega}\Big)^2\Big\}}  A\Big(\frac{T}{T_c}\Big) \nonumber \\
& \times & \cos(2\pi \frac{f_{a}}{B}) \cos(2\pi \frac{f_d}{B}) \Big],
\end{eqnarray}
where $ f_a = (f^+ + f^-)/2 $ and $ f_d = (f^+ - f^-)/2 $.  
It clearly shows that the total conductivity produces beating patterns in 
the amplitude of the SdH oscillations.
In Fig. 1, we compare the analytical result of the total conductivity
with the exact numerical results obtained from Eq. (\ref{exact}).
The beating pattern obtained from the analytical expression is in
excellent agreement with the numerical results, particularly the locations
of the nodes.
For Fig. 1, the following parameters are used:
$\alpha = 10^{-11} $ eV-m, $\Gamma_0 = 0.02$ meV,
electron density $n_e = 3 \times 10^{15}$ /m${}^2$, 
electron effective mass $m^{\ast} = 0.05 m_0 $ 
with $m_0$ is the free electron mass, and temperature $ T = 1 $ K.

\begin{figure}[t]
\begin{center}\leavevmode
\includegraphics[width=98mm]{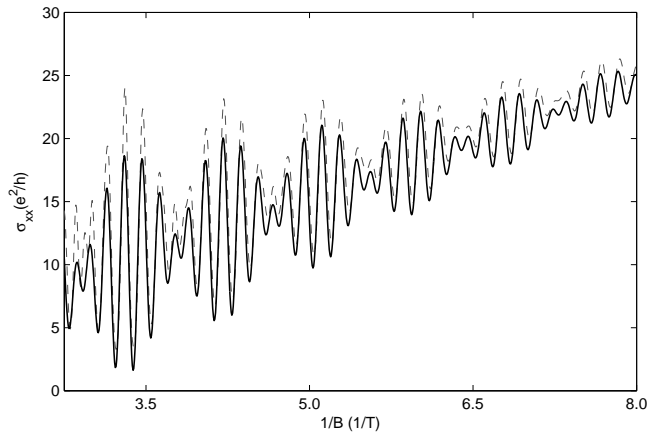}
\caption{Plots of the exact (dashed) and analytical (solid) expression of
the total conductivities vs  inverse magnetic field $B$.}
\label{Fig1}
\end{center}
\end{figure}

The number of oscillations between two successive nodes is
\begin{eqnarray}
N_{osc} =  f_a\Delta(\frac{1}{B}) =
\frac{m^*}{\hbar e} 
\Big(E_F + \frac{E_{\alpha}}{2}\Big)\Big(\frac{1}{B_{j+1}}-\frac{1}{B_{j}}\Big),
\end{eqnarray}
where $ B_j$ is the magnetic field corresponding to the $j$-th node. 

The last cosine term, $ \cos( 2\pi f_d/B) $ is not periodic in $1/B$
because the frequency difference, 
$f_d = \frac{m^*}{\hbar e} \sqrt{E_0^2 + E_{\alpha}E_{F}} $,
itself depends on the magnetic field $B$. 
The non-periodic behavior of the beating patterns observed in the 
experiments is due to the magnetic field dependence of the term $f_d$.

At the node positions $ B = B_j$, we have the following condition:
$\cos(2\pi f_d/B)|_{B=B_j}=0 $, which gives us 
\begin{equation}
\sqrt{4 E_0^2 + \Delta_s^2} = \hbar\omega_j (j+\frac{1}{2}) \label{beat},
\end{equation}
where $ \Delta_s = 2 k_{_F} \alpha $ is the zero-field spin splitting energy, 
$j=1,2,3..$ are the $j$-th beat node and $\omega_j = eB_j/m^* $. 
The zero-field spin splitting energy can be easily evaluated by knowing the
magnetic field corresponding to any beat node.
The above equation can be re-written as
\begin{equation}
B_j = \frac{2 m^*}{e \hbar} 
\frac{\Delta_s}{\sqrt{(2j + 1)^2 - (g -2)^2}}. 
\end{equation}
The above equation is exactly the same as given in Ref. \cite{other}, 
which has been obtained by using
the self-consistency Born approximation.

\subsection{Comparison with Experiment}
In this section we would like to test our analytical expressions by
calculating the positions of the nodes, the zero-field spin splitting 
energy and the number of oscillations between two nodes and then 
comparing with the experimental observations. 
To reproduce the experimental observations of the beating patterns in
the SdH oscillations \cite{dutta}, we plot the resistivity as a function 
of the magnetic field in Fig. 2.
For Fig. 2, we have taken the following parameters as used in the 
experiment \cite{dutta} for sample A: $\alpha = 3.76 \times 10^{-12} $ eV-m, 
$\Gamma_0 = 1.5$ meV,
electron density $n_e = 1.75 \times 10^{16}$ /m${}^2$, 
electron effective mass $m^{\ast} = 0.046m_0 $ with $m_0$ is the 
free electron mass and temperature $ T = 1.5 $ K.
\begin{figure}[t]
\begin{center}\leavevmode
\includegraphics[width=98mm]{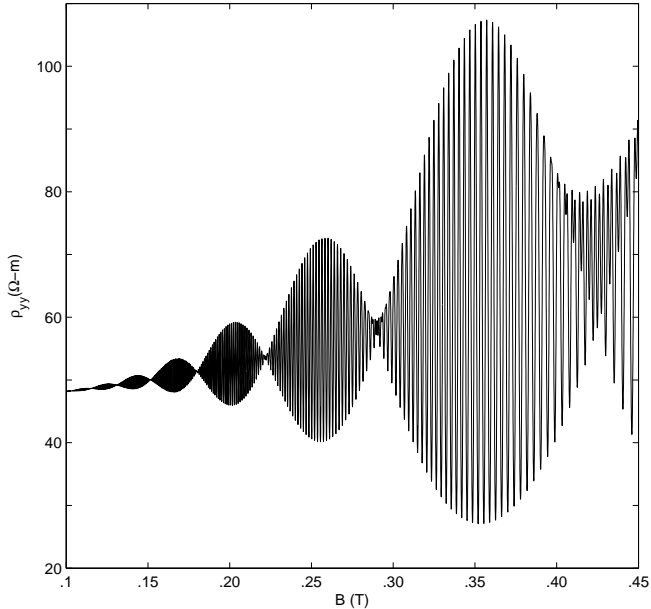}
\caption{Plots of the total resistivity vs magnetic
field $B$.}
\label{Fig2}
\end{center}
\end{figure}
It is interesting to note that the locations of the beat nodes and the
number of oscillations between two successive nodes match very well
with the experimental observations \cite{dutta}.

We would like to determine the zero-field spin splitting energy and hence 
the Rashba SOI strength by using the positions of the beat nodes for 
three different samples A, B, and C used in Ref. \cite{dutta}.
In Ref. \cite{dutta}, the following parameters are used: 
$m^{\ast} = 0.046 m_0 $ and $n_e = 1.75 \times 10^{16} $ m${}^{-2}$, 
$ 1.65 \times 10^{16} $ m${}^{-2}$, $ 1.46 \times 10^{16} $m${}^{-2}$ 
for the A, B, and C samples, respectively. 
Using Eq. (\ref{beat}),
we determine the zero-field spin splitting energy and the SOI strength 
for the first six nodes as given in the Table I,
\begin{table}[ht]
\centering
\begin{tabular}{|c |c |c |c |}
\hline\hline
node  & Beat points &$\Delta_s^A$& $ \alpha $ \\
&($ B $) in Tesla &  meV  & $(10^{-12})$ eV-m \\
\hline
1 & 0.873 & 2.44 & 3.69 \\
\hline
2 & 0.460 & 2.63 & 3.98 \\
\hline
3 & 0.291 & 2.43 & 3.68 \\
\hline
4 & 0.227 & 2.49  & 3.76 \\
\hline
5 & 0.183 & 2.47  & 3.74 \\
\hline
6 & 0.153 & 2.45  & 3.71 \\
\hline
\end{tabular}
\caption{Sample A: Zero-field spin splitting energy and the 
Rashba SOI strength $\alpha$ for different positions of the
beat nodes.} 
\end{table}
The average values of the zero-field spin splitting energy at the 
Fermi level is $\Delta_s^{A} = 2.49$ meV and the Rashba SOI is 
$3.76 \times 10^{-12}$ eV-m. Our results are nearly same as obtained in
Ref. \cite{dutta}.
Similarly for the sample B, we obtain the following results given in 
the Table II:

\begin{table}[ht]
\centering
\begin{tabular}{|c |c |c |c |}
\hline\hline
node  & Beat points &$\Delta_s^B$& $ \alpha$ \\
&($B$) in Tesla & meV & $(10^{-12})$ eV-m\\
\hline
2 & 0.294  & 2.69 & 4.19 \\
\hline
3 & 0.200  & 2.69 & 4.19 \\
\hline
4 & 0.152 & 2.67 & 4.15 \\
\hline
5 & 0.128  & 2.77 & 4.31 \\
\hline
6 & 0.103  & 2.67 & 4.15 \\
\hline
\end{tabular}
\caption{Sample B: Zero-field spin splitting energy and the 
Rashba SOI strength $\alpha$ for different positions 
of the beat nodes.}
\end{table}

The average value of the zero-field spin splitting energy 
is $\Delta_s^{B} = 2.69$ meV and the Rashba SOI strength is
$\alpha = 4.19 \times 10^{-12}$ eV-m.
Similarly, for sample C, we obtain 
the average value of the zero-field spin splitting 
energy is $\Delta_s^{C} = 1.76$ meV 
and the Rashba SOI strength is $ \alpha = 2.91 $ eV-m. 
These results are in excellent agreement with the result obtained  
in Ref. \cite{dutta}.

Now we consider another experiment where the SOI strength has been measured 
\cite{luo}. By knowing the positions of the two successive node points, 
we can also calculate the SOI strength. Using the parameters used 
in Ref. \cite{luo}, we obtain the Rashba SOI strength  
$\alpha = 0.9 \times 10^{-9}$ eV-cm which is exactly the same as 
obtained in Ref. \cite{luo}.                                                            

The number of oscillations between two successive nodes $j=1$ and $j=2$
counted from Ref. \cite{dutta} is 36 [see figure of Ref. \cite{dutta}].
Using the parameters used in the experiment \cite{dutta}, we obtain
$ N_{osc} = 37 $ which exactly matches with the experimental observations.
We consider another experiment \cite{tech} where the SOI strength was
varied by varying the gate voltage. For the gate voltages $V_g=0.3$ 
and 1.5 Volt, the calculated number of oscillations between two nodes 
are $ N_{osc}=27$ and 30, respectively. These numbers are the same 
with the direct observations.

\section{CONCLUSION}
In this work, the beating patterns in the SdH oscillations in a 2DEG with 
the Rashba SOI is revisited.
We have derived the DOS for spin-up and spin-down electrons 
in a 2DEG with the Rashba SOI in presence of a magnetic field analytically.
The analytical expressions of the DOS will be very useful to calculate 
other properties, like magnetization, susceptibility etc, of a 2DEG with 
the SOI analytically. 
We have provided analytical expressions of the magnetoconductivities for 
spin-up and spin-down electrons, which oscillate with two closely different frequencies.
The frequencies of the conductivity oscillations depend on the electron density,
the SOI strength and also the external magnetic field.
We have used the most conventional approach to get the
simple equation which determines the zero-field spin splitting energy 
by knowing the magnetic field corresponds to any beat node.
The number of oscillations in any beat can be easily found from our expression by knowing
the two beat points. 
The calculated number of oscillations in a beat exactly matches with the experimental
findings. The non-periodic beating pattern is due to the magnetic field dependence
of the frequency difference between spin-up and spin-down electrons.

%\begin{appendix}
%\section{}
%The self energy can be written as
%\begin{eqnarray}
%\Sigma^-(E) & = & \Gamma_0^2 \sum_{s} \frac{1}{E-E^{\pm}_{s} - \Sigma^-(E)}
%\nonumber\\ 
%& = & \Gamma_0^2 \sum_{s}f^{\pm}(s),
%\end{eqnarray}
%where
%\begin{equation}
%f^{\pm}(s)=\frac{1}{E-\Sigma^-(E) - 
%s \hbar \omega_0 \mp \sqrt{E_0^2 + s \hbar \omega_0E_{\alpha}}},     
%\end{equation}
%$\Gamma_0$ is the broadening of the Landau levels due to impurities and 
%$E_i$ is the sub-band energy. By determining 
%the imaginary part of self-energy we can get density of states through
%\begin{equation}
% D(E)=\Im\left[\frac{\Sigma^-(E)}{\pi^2 l_0^2 \Gamma_0^2}\right] \label{density}.
%\end{equation}
%To find the above sum we use the following residue theorem:\\
%$\sum_{s=-\infty}^{\infty}f^{\pm}(s)=-$ $\{$sum of all residues of 
%$[\pi \cot(\pi s)f^{\pm}(s)]$
%at poles of $f^{\pm}(s)$ $\}$.

%Finally the DOS can be obtained for spin up and spin down branches as
%\begin{eqnarray}
%D^{\pm}(E)&=&\frac{1}{\pi l_0^2}\frac{1}{\hbar\omega_0}
%\Big[1+2\exp\Big\{-2(\frac{\pi\Gamma_0}
%{\hbar\omega_0})^2\Big\}\nonumber\\&\times&
%\cos \Big\{\frac{2\pi}{\hbar\omega_0}
%\Big(E+\frac{E_{\alpha}}{2}\mp\sqrt{E_0^2+E_{\alpha}E}\Big)\Big\}\Big]\nonumber\\
%\end{eqnarray}

%\end{appendix}

\section{acknowledgement}
This work is financially supported by the CSIR, Govt. of India under the grant
CSIR-JRF-09/092(0687) 2009/EMR-I, F-O746.


\begin{thebibliography}{55}

\bibitem{rmp1}
I. Zutic, J. Fabian, and S. Das Sarma, Rev. Mod. Phys. {\bf 76},
323 (2004).


\bibitem{rmp2}
F. Fabian, A. Matos-Abiague, C. Ertler, P. Stano, and
I. Zutic, Acta Physica Slovaca {\bf 57}, 565 (2007).


\bibitem{book}
S. Bandyopadhyay and M. Cahay, Introduction to Spintronics
(CRC Press-2008).

\bibitem{das1}
S. Datta and B. Das, 
Appl. Phys. Lett. {\bf 56}, 665 (1990).

\bibitem{bandyo}
S. Bandyopadhyay and M. Cahay, Appl. Phys. Lett. {\bf 85}, 1814 (2004).

\bibitem{mit}
E. Tutuc, E. P. De Poortere, S. J. Papadakis, and M. Shayegan,
Phys. Rev. Lett. {\bf 86}, 2858 (2001);
S. J. Papadakis et al., Science {\bf 283}, 2056 (1999).

\bibitem{ballistic}
J. P. Lu, J. B. Yau, S. P. Shukla, M. Shayegan, L. Wisinger,
U. Rossler, and R. Winkler, Phys. Rev. Lett. {\bf 81}, 1282 (1998).

\bibitem{galvanic}
S. D. Ganichev, E. L. Ivchenko, V. V. Bel'kov, S. A. Tarasenko,
M. Sollinger, D. Weiss, W. Wegscheider, and W. Prett, Nature (London)
{\bf 417}, 153 (2002).

\bibitem{perel}
M. I. Dyakonov and V. I. Perel, JETP Lett. {\bf 13}, 467 (1971).

\bibitem{she}
S. Murakami, N. Nagaosa, and S. C. Zhang, Science {\bf 301}, 1348 (2003).

\bibitem{dress}
G. Dresselhaus, Phys. Rev. {\bf 100}, 580 (1955).

\bibitem{rashba}
E. I. Rashba and V. I. Sheka, 
Dokl. Akad. Nauk SSSR {\bf 2}, 162 (1959);
E. I. Rashba, Sov. Phys. Solid State {\bf 2}, 1109 (1960).

\bibitem{lommer}
G. Lommer, F. Malcher, and V. Rossler, 
Phys. Rev. Lett. {\bf 60}, 728 (1988); 
Phys. Rev. B {\bf 32}, 6965 (1985).


\bibitem{bassani}
E. A. de Andrad e Silva, G. C. La Rocca, and F. Bassani,
Phys. Rev. B {\bf 50}, 8523 (1994).

\bibitem{stormer}
H. L. Stormer, Z. Schlesinger, A. Chang, D. C. Tsui,
A. C. Gossard, and W. Weigmann, 
Phys. Rev. Lett. {\bf 51}, 126 (1983).


\bibitem{stein}
D. Stein, K. v. Klitzing, and G. Weimann, 
Phys. Rev. Lett. {\bf 51}, 130 (1983).

\bibitem{rashba1}
Y. A. Bychkov and E. I. Rashba, 
J. Phys. C {\bf 17}, 6039 (1984). 


\bibitem{luo}
J. Luo, H. Munekata, F. F. Fang, and P. J. Stiles, 
Phys. Rev. B {\bf 38}, 10142 (1988); {\bf 41}, 7685 (1990).

\bibitem{dutta}
B. Das, D. C. Miller, S. Datta, R. Reifenberger, W. P. Hong,
P. K. Bhattachariya, J. Sing, and M. Jaffe, 
Phys. Rev. B {\bf 39}, 1411 (1989).

\bibitem{tech}
J. Nitta, T. Akazaki, H. Takayanagi, and T. Enoki,
Phys. Rev. Lett. {\bf 78}, 1335 (1997).

\bibitem{tech1}
G. Engels, J. Lange, Th. Schpers, and H. Luth,
Phys. Rev. B {\bf 55}, R1958 (1997).

\bibitem{tech2}
J. P. Heida, B. J. van Wees, J. J. Kuipers, T. M. Klapwijk,
and G. Borghs,
Phys. Rev. B {\bf 57}, 11911 (1998).


\bibitem{das}
B. Das, S. Datta, and R. Reifenberger, Phys. Rev. B {\bf 41}, 8278 (1990)

\bibitem{wang}
X. F. Wang and P. Vasilopoulos, Phys. Rev. B {\bf 67}, 085313 (2003).

\bibitem{other}
S. G. Novokshonov and A. G. Groshev, Phys. Rev. B {\bf 74}, 
245333 (2006).


\bibitem{ando}
T. Ando, J. Phys. Soc. Jp. {\bf 37}, 1233 (1974);
T. Ando, A. B. Fowler, and F. Stern, 
Rev. Mod. Phys. {\bf 54}, 437 (1982).

\bibitem{ger}
C. Zhang and R. R. Gerhadtz, Phys. Rev. B {\bf 41}, 12850 (1990).

\bibitem{kubo}
P. Vasilopoulos and C. M. Van Vilet, J. Math. Phys.
{\bf 25}, 1391 (1984).

%\bibitem{peeters}
%X. F. Wang, P. Vasilopoulos, and F. M. Peeters, 
%Phys. Rev. B {\bf 71}, 125301 (2005).



\end{thebibliography}
\end {document}